\documentclass[a4paper,fleqn]{cas-sc}

\usepackage{graphicx} 
\usepackage[authoryear]{natbib}
\usepackage{threeparttable}
\usepackage{lineno}

\begin{document}
\let\WriteBookmarks\relax
\def\floatpagepagefraction{1}
\def\textpagefraction{.001}

\shorttitle{Meteor clusters}
\shortauthors{P. Koten et~al.}

\title [mode = title]{Meteor clusters: tracing meteoroid fragmentation in near-Earth space}                      

\author[1]{P. Koten}[type=editor,
                        orcid=0000-0002-7079-4489]
\cormark[1]
\ead{pavel.koten@asu.cas.cz}

\credit{Conceptualization of this study, Methodology, Software}

\affiliation[1]{organization={Astronomical Institute, Czech Academy of Sciences},
                addressline={Fri\v{c}ova 298}, 
                city={Ond\v{r}ejov},
                postcode={25165}, 
                country={Czech Republic}}

\author[1]{D. \v{C}apek}

\author[2]{J. T\'{o}th}

\affiliation[2]{organization={Faculty of Mathematics, Physics and Informatics, Comenius University},
                addressline={Mlynsk\'{a} Dolina}, 
                city={Bratislava},
                postcode={84248}, 
                country={Slovakia}}

\author[3]{J. Vaubaillon}

\affiliation[3]{organization={Institut de M\'{e}canique C\'{e}leste et de Calcul des \'{E}ph\'{e}m\'{e}rides},
                addressline={77 avenue Denfert-Rochereau}, 
                city={Paris},
                postcode={75014}, 
                country={France}}
                
\author[3]{A. Ashimbekova}                

\author[3]{S. Anghel}             

\author[4]{J. Watanabe}

\affiliation[4]{organization={National Astronomical Observatory},
                addressline={2-21-2 Osawa}, 
                city={Mitaka, Tokyo},
                postcode={181-8588}, 
                country={Japan}}

\author[2]{T. V\"{o}r\"{o}s}                
                
\cortext[cor1]{Corresponding author}


\begin{abstract}
Meteor clusters are typically defined as groups of meteors that appear close together in both space and time. To date, only a handful of such events have been recorded instrumentally and analysed in detail. In many documented cases, thermal stress has been identified as the most likely cause of meteoroid fragmentation near Earth. This paper documents two further cases and provides a summary of all currently known clusters. The two clusters that were recorded over Hawaii Island in 2023 and 2024 represent two distinct scenarios. The 2024 meteor cluster was characterised by a dominant mass body and, with the fragments arranged along the antisolar direction according to their mass. Such cases enable us to reliably determine the age of the cluster and identify the most likely formation scenario. This cluster was around three days old, and the thermal stress was the most likely mechanism of its formation. The 2023 cluster was not such a case. It does not contain a mass dominant body, nor are its fragments arranged by their mass. Therefore, it was only possible to estimate its age to be no more than four days. Furthermore, other potential formation mechanisms besides thermal stress cannot be ruled out. This fact was observed in all analysed clusters. All clusters known up to date were formed in close proximity to Earth. The volume of a cluster increases with its age. This means that older clusters, formed by the fragmentation far away from Earth may remain undetected, as their fragments are also dispersed too widely to be observed by local experiment. However, global networks can detect such dispersed clusters.
\end{abstract}


\maketitle

\section{Introduction}

A meteor cluster is a phenomenon that occurs when the meteoroids close in both space and time and of the same origin enter the Earth's atmosphere. Numerous visual and telescopic observers reported pairs and/or groups of meteors. However, as these observations are subjective, it is the instrumental records that provide a more comprehensive understanding of such events. Exceptional visual observations are those events that were witnessed by a large number of people. One example is the famous meteoritic procession, that occurred on 9 February 1913 \citep{Chant1913}. Hundreds or even thousands of people witnessed it along a 15,000-kilometre tract stretching from the Canada via the United States and the Bermuda Islands to the north-eastern coast of Brazil. The procession consisted of at least ten groups, each containing tens of meteors. \cite{Beech2018} suggested that the event was caused by an object that had been temporarily captured in an Earth orbit and disintegrated before its fragments entered the atmosphere. 

The first instrumental record of the meteor cluster was obtained on 12 August 1977 with double-station LLLTV cameras. It consists of three Perseid meteors that arrived on nearly parallel trajectories within 1.3~seconds. \cite{Hapgood1981} deduced that the parent meteoroid broke up at least 1700~km from the Earth's surface. Another event was recorded on 18 October 1985, using an image intensifier video system. A cluster of five almost simultaneous meteors on parallel trajectories was detected, occurring within a fraction of a second. Each fragment itself was composed of at least four smaller meteoroids \citep{Piers1993}.

Three large meteor clusters were recorded during the Leonid meteor storms at the turn of the century. The first of these occurred in 1997, when 100-150 Leonid meteors appeared in just two seconds \citep{Kinoshita1999}. Two more clusters were observed during a period of high Leonid activity in 2001 \citep{Watanabe2002}. One of these consisted of at least 15 meteors, which appeared within four seconds. The second contained 38 meteors recorded over a period of two seconds. \cite{Watanabe2003} concluded that the fragmentation occurred close to the Earth, because all the fragments of all three clusters were distributed in a region of just a few hundred kilometres. The most likely point at which this occurred was found to be perihelion of the orbit, which is due to the shape of the orbit just six days before the encounter with our planet.

The double-station observation of the cluster belonging to the September epsilon Perseid meteor shower allowed the first detailed analysis of such an event. The fireball, captured by the fireball network, was accompanied by nine faint meteors that were recorded by the video cameras over a period of 1.5 seconds. Due to the proximity and size of the fragments it was concluded that the fragmentation occurred no more than two or three days before entering the atmosphere \citep{Koten2017}. This cluster was analysed using a method that was developed for this purpose. It was determined that the age of the cluster was 2.28$\pm$0.44 days. The ejection velocities of the individual fragments range from 0.13 to 0.77 m.s$^{-1}$. Exfoliation from the surface of the main body due to the thermal stress was found to be the most likely mechanism of cluster formation \citep{Capek2022}.

Several other meteor clusters have been either observed or discovered in the archives in recent years. Among them was a huge meteor cluster that was observed above the Baltic Sea and southern Scandinavia in 2022 \citep{Koten2023}, and a cluster recorded during the $\tau$-Herculid meteor cluster outburst in the same year \citep{Vaubaillon2023, Koten2025}. Archive searches revealed records of two clusters: a compact one observed in Brazil in 2019 \cite{Koten2024b}, and a Leonid meteor shower cluster observed in Spain in 2002 \citep{Koten2025b}. Three additional clusters were recorded by a unique Subaru-Asahi StarCam located in Hawaii \citep{Abe2026, Watanabe2025}.

Although the number of recorded meteor clusters has increased in recent years, they remain a rare phenomenon. Based on statistics of the number of instrumentally recorded meteors, \cite{Vaubaillon2023} estimate that a meteor cluster occurs once in every million meteor observations. Understanding of the processes that lead to the fragmentation of the meteoroids in the interplanetary space will reveal the true nature of this fascinating phenomenon and also provide astronomers with an insight into the life expectancy of the meteoroids in the space.

In this paper, we present a detailed analysis of two additional clusters and summarise all the available cases. We also begin to identify the common patterns among them. The structure of the paper is as follows: Firstly, Section~\ref{nomenclature} proposes some definitions regarding the meteor clusters.  Section~\ref{Hawaii_clusters} provides a detailed description of two meteor clusters that were observed over Hawaii in 2023 and 2024, respectively. Section~\ref{cluster_summary} provides an overview of all clusters analysed in detail, and Section~\ref{conclusions} summarises the current knowledge on the clusters.

\section{Meteor clusters nomenclature}
\label{nomenclature}

In recent papers, meteor clusters were mostly designated according to their place of observation or their affiliation with a meteor shower. Until now, no systematic approach has been adopted. The rule for naming meteor clusters was discussed with members of the IAU Commission F1 Working Group on Meteor Shower Nomenclature. The working group proposes the following convention: a cluster whose first member was observed on DD-MM-YYYY (where DD represents the day, MM the month, and YYYY the year of observation) would be designated MC/YYYY-MM-DD. If more than one cluster is observed on the same date, a small letter starting with 'b' will be added.  We start to using the proposed convention in the following summary of the clusters. According to this rule, the cluster observed on 19 November 2002, for example, would be labelled 'MC/2002-11-19'.\footnote{This rule is still a proposal to be discussed at the next WG business meeting.}

As the definition of the meteor cluster does not exist, we also provide some hints on how to identify such events. It is usually supposed that the cluster is a group of meteoroids that are close together in space and time. At this moment, we would like to avoid setting precise limits of temporal and spatial proximity. Instead, we prefer that two conditions be satisfied for a group to be accepted as a cluster: Firstly, the activity of the cluster meteors must be higher than expected activity of the meteor shower or of the sporadic background. Secondly, the common dynamical and/or physical origin of the meteoroids of the group must be demonstrated.

A statistical test is the best way to demonstrate that the group of meteors satisfies the first condition. A Monte Carlo simulation can be performed using the knowledge of the shower or sporadic activity to show whether a group of meteors with a measured time distribution is the result of coincidence or it is a real cluster (at a chosen level of confidence). For example, this statistical test was employed by \cite{Koten2021} when searching for the pairs and groups among the Geminid meteors.

Regarding the second condition, when analysing recent clusters, we have measured the positions of the meteoroids relative to each other and then attempted to determine their common origin. This method has been introduced in \cite{Capek2022}.

\section{Two clusters observed from Hawaii islands}
\label{Hawaii_clusters}

The Subaru-Asahi StarCam, a high-sensitivity live camera installed at the summit of Mauna Kea in Hawaii, has already recorded three meteor clusters. \cite{Abe2026} have analysed the first event that occurred in 2021. Only basic data has been published for the other two events that were recorded in 2023 and 2024. \cite{Watanabe2025} reported the appearance of the clusters and provided basic characterisation. They also discussed the possible shower membership and estimated probability of detection of the clusters by the camera. 

This paper provides additional information on both clusters. The video records of clusters have been processed, and the positions of the meteors have been measured manually using Fishscan software.  \citep{Borovicka2022}. The 2023 cluster was recorded only from a single station. It means that a number of solutions are consistent with a measured direction and angular velocity of each individual meteor. Fortunately, as demonstrated in the case of the Brazil cluster \citep{Koten2024b}, each cluster can be treated as a small meteor shower with all the meteors originating from the same radiant. This condition enables the identification of the radiant of the cluster. In the case of the 2024 cluster, two brightest meteors were also recorded by the AMOS cameras at Mauna Kea and Haleakala observatories. The measured radiant was used also for calculation of other, single station meteors. With known radiant it is possible to determine the atmospheric trajectories of all cluster meteors. In both cases, the same numbering was used to be consistent with the original report. In the case of the second cluster, two more meteors were found and analysed.

\subsection{Instrumentation}

Two clusters near the Hawaiian islands were recorded  using two types of instrument in 2023 and 2024. The Subaru–Asahi StarCam, a high sensitivity live camera installed at the summit of Maunakea, Hawaii, on the site of the Subaru Telescope, commenced live streaming of the night sky \citep{Tanaka2025}. Initially planned for public outreach and dissemination, its high clear sky rate, excellent atmospheric transparency, and the high sensitivity of its advanced camera have led to the detection of numerous meteors. Consequently, it has garnered significant attention from online viewers, capturing not only the activity of various meteor showers but also cluster phenomena. Following the initial detection of a cluster event in 2021 \citep{Abe2026}, the StarCam has successfully detected cluster events on 20 April 2023 and 11 July 2024. 

The pixel scale of the streaming video, which was approximately 0.0095$^{\circ}$ for the 2023 cluster and 0.018$^{\circ}$ for the 2024 cluster (3840 × 2160 and 1920 × 1080 pixels, respectively). The different pixel scales are due to different cameras used at the epoch of the clusters’ apparition. While on 2023 April 20 was used the camera Sony Alpha 7S III with a lens of Sony 24mm F1.4 GM with streaming HD( 1080p), on 2024 July 11 the camera Sony FX3 with a lens of Sony 24mm F1.4 GM with streaming 4K (2160p) was used. The frame rate was 30 fps in both cases. The detailed information is described by \cite{Tanaka2025}. 

The All-Sky Meteor Orbit System (AMOS)\footnote{\url{https://amos.uniba.sk}} is an automated meteor all-sky and spectral observatory system consisting of 17 stations of the AMOS global meteor network covering the north and South hemispheres and different longitudes. It continuously monitors meteor activity every night. (\cite{Toth2025}). The network locations include Slovakia, the Canary Islands, Chile, Hawaii, Australia, South Africa, and Arizona. Each station has a digital video CCD camera for all-sky observation of 1600x1200 pixels and 20 frames per second, covering a wide range of meteor magnitudes, typically between +4.0 and -3.0 (\cite{Toth2015}) and even fireballs. AMOS's primary goal is to calculate precise trajectories and orbits of meteors and correlate them with their spectral properties studied by the AMOS-Spec cameras to improve the knowledge of weak meteor showers and their parent body characterization, or to provide comprehensive studies of meteoroid dynamics, physical properties, and composition \citep{Matlovic2019, Matlovic2022}. The network has successfully integrated software algorithms to determine the atmospheric trajectory, heliocentric orbits, and physical parameters of meteoroids based on standard methods in meteor astronomy \citep{Toth2025}. These cameras are part of an integrated system designed for advanced meteor detection and analysis, which contributes to research on the origins of meteoroids and the potential recovery of meteorites.

\subsection{Cluster 20. 4. 2023}

\begin{figure}[htbp]
  \centering
  \includegraphics[width=0.48\textwidth]{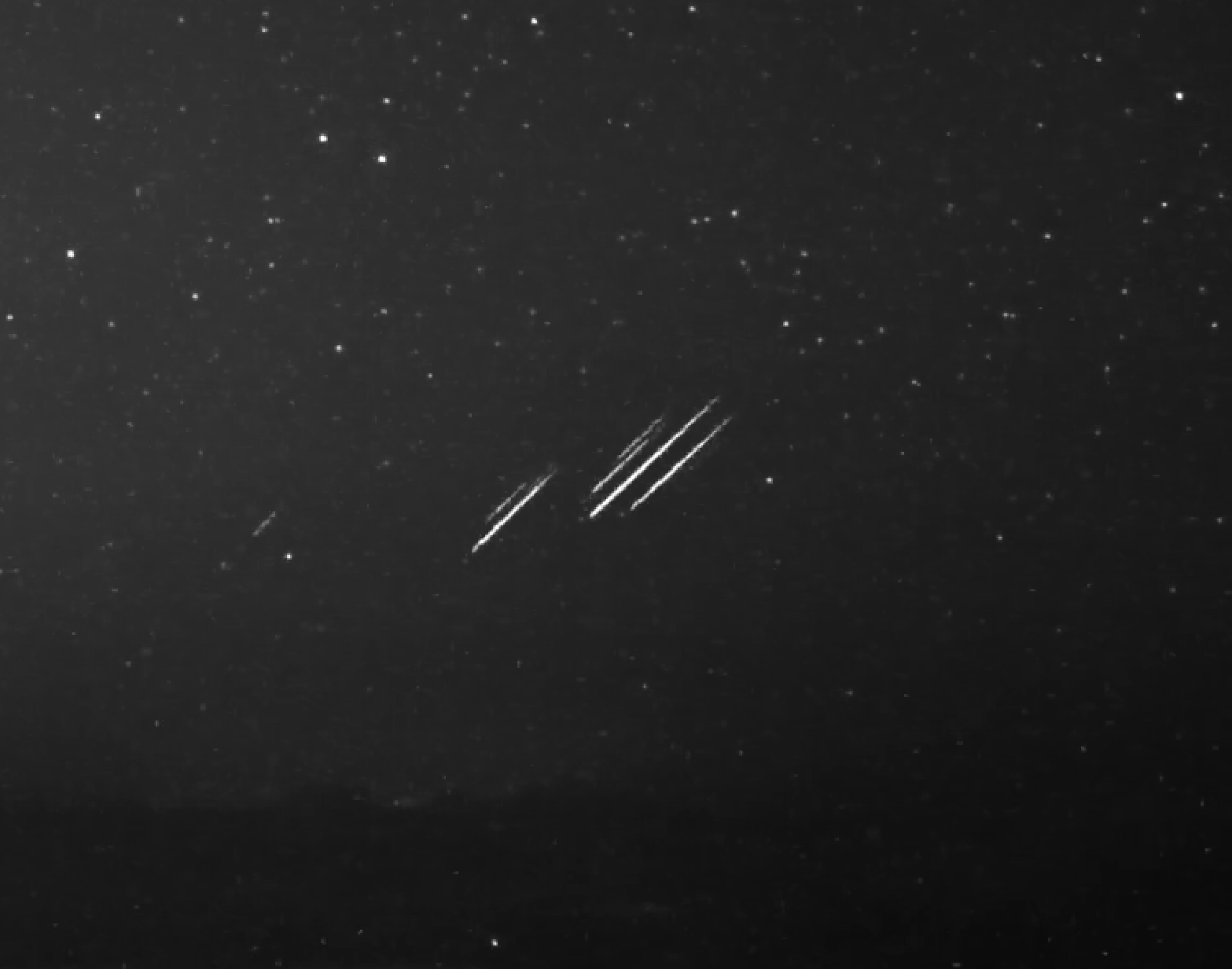}
  \includegraphics[width=0.48\textwidth]{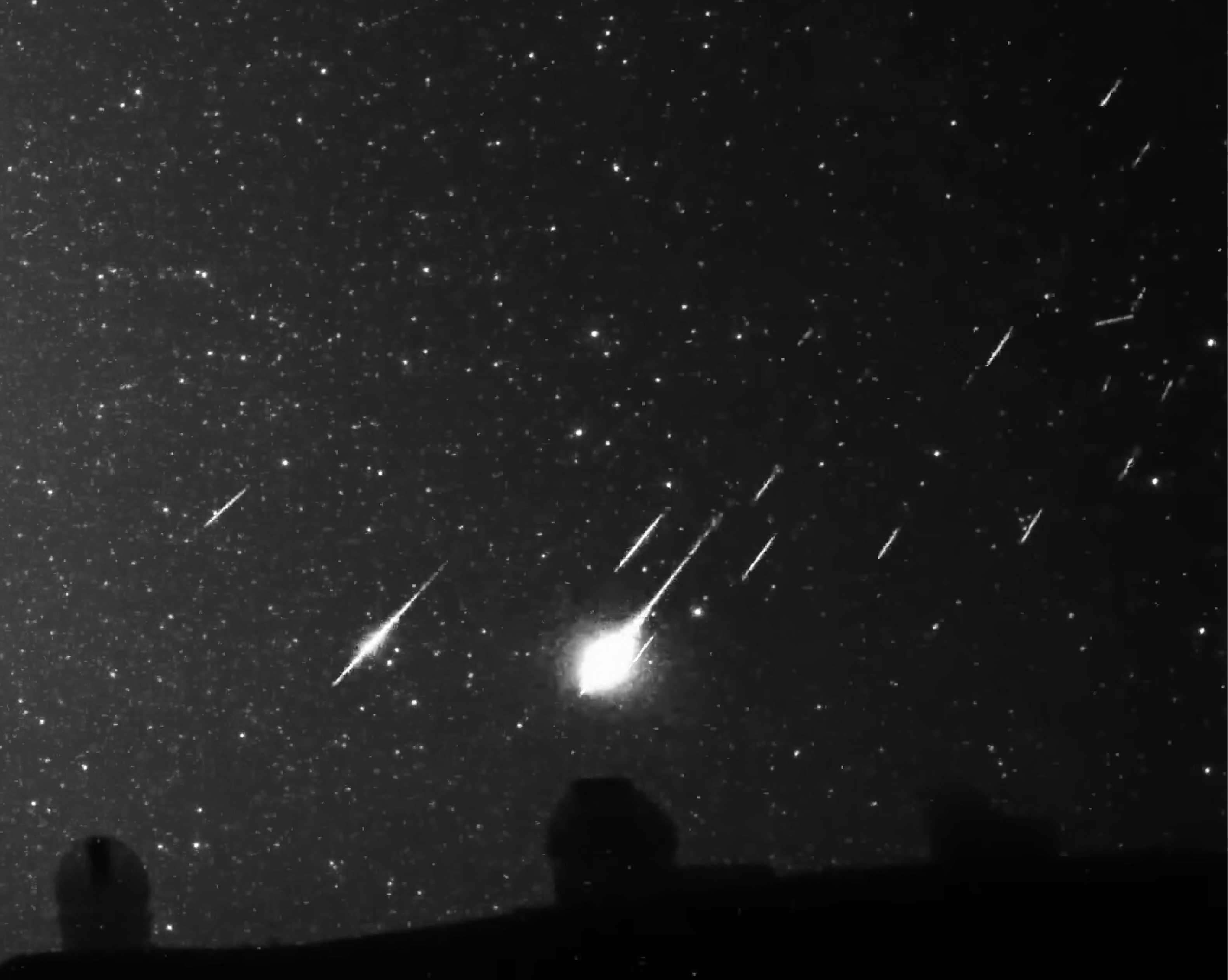}
  \caption{A composite image of seven meteors of the 2023 cluster (left panel) was created by the stacking of the video frames recorded by the Subaru-Asahi StarCam. The right panel shows 16 meteors of the 2024 cluster as recorded by the same camera.}
  \label{fig_composite}
\end{figure}

This cluster was observed by the Subaru-Asahi StarCam on 20th April 2023 at 11:50:25~UT. It was comprised of seven meteors, with the appearance of the last one less than three seconds after the first one. The composite image of the cluster is shown in the left panel of Figure~\ref{fig_composite}.

As is evident from the image, that this was not an event of a mass-dominant main body. The photometric measurements showed that the heaviest meteoroid contributed to the total mass of the cluster by only 58\%. 

\begin{figure*}[htbp]
  \centering
  \includegraphics[width=0.9\textwidth]{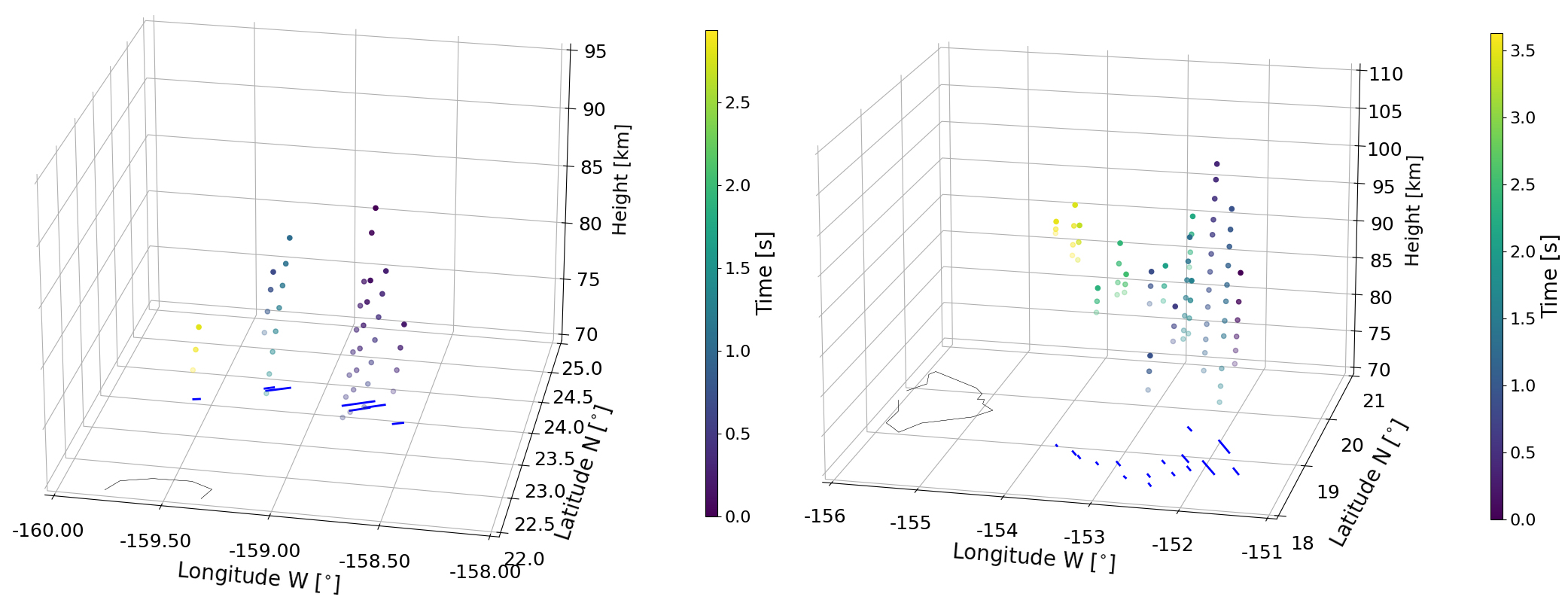}
  \caption{Atmospheric trajectories of the 2023 (left panel with Kaua's island) and 2024 (right panel with Hawaii island) cluster. Time elapsed from the appearance of the first meteor is colour coded. Although the trajectories appear very steep in both cases, the angles to the vertical are approximately 42$^{\circ}$ and 50$^{\circ}$ respectively.}
  \label{fig_3D_maps}
\end{figure*}

\cite{Watanabe2025} explored the potential shower membership of the cluster. One of the candidates was the R-Lyrid meteor shower\footnote{653 RLY according to the IAU Meteor Data Center database}. When atmospheric trajectories were calculated, it turned out that for the majority of meteors in the cluster, the radiant R Lyrid represents a suitable solution. Consequently, the atmospheric trajectories were determined using this radiant ($\alpha_{G} = 280.4^{\circ}$, $\delta_{G} = 47.6^{\circ}$, $v_{G} = 40.1~km.s^{-1}$). The maximum activity of the meteor shower is at $\lambda_{\odot} = 32^{\circ}$ \citep{Jenniskens2016}. The cluster has been observed at $\lambda_{\odot} = 29.83029^{\circ}$, i.e. close to the maximum. The trajectories are plotted in the left panel of Figure~\ref{fig_3D_maps}. The zenith distance of the radiant was 42.3$^{\circ}$. According to the proposed rule, the cluster should be labelled MC/2023-04-20.

The cluster occurred approximately 200 km north of the Kaua'i island. The distance from the camera located on the Hawaii island was about 550 km. The figure also demonstrates that the cluster was gradually shifting from the east to the west. The first meteor was the brightest (in the image the second from the right), rapidly succeeded by five meteors within one second. The last one, positioned on the left, appeared with a delay of 2.8 seconds. It was the faintest meteor of the cluster.

To investigate the cluster origin, it is necessary to determine the mutual positions and masses of the meteoroids. Therefore, the positions of the meteoroids outside the atmosphere are calculated. The observed minimum distance between the fragments was 13.3~km, and the maximum was 94~km. In general, the fragment 7 (the last one) was the most dispersed with an average distance to other fragments measuring 71~km. The total volume of the cluster was found to be approximately 245.9 thousand km$^{3}$.

\subsection{Cluster 11. 7. 2024}

The cluster was observed by the same camera on 11 July 2024 at 14:05:33~UT. It was the third cluster captured by the Subaru-Asahi StarCam in three years. The cluster consisted of 15 meteors that were observed within 3.8~second. A bright fireball was among them. The composite image is shown in the right panel of Figure~\ref{fig_composite}.

In this case, two meteors were also recorded by the AMOS cameras and they can be treated as the double station meteors. The only limitation was unfavourable angle of the planes, which was about $8.6^{\circ}$ in the case of the brightest meteor (meteor number 2). Despite this fact, a reliable trajectory was obtained. The trajectory calculation was more complicated for the second brightest meteor (number 5) because it was only recorded in a few video frames. However,  a reasonable solution was also found for this meteor. The atmospheric trajectories of both meteors are summarized in Table~\ref{tab_20240711_trajectories}. With inclination of about $127^{\circ}$ neither meteor shower nor NEA object was found to exist close to the cluster's heliocentric orbit. The cluster should be labelled MC/2024-07-11 in reference to its date of observation.

\begin{table*}
\begin{threeparttable}
\caption{Atmospheric trajectories of double station meteors of the 2024 cluster.}
\label{tab_20240711_trajectories}
\centering          
\begin{tabular}{l c c c c c c c c c}
\hline
no.    & $\alpha_{G}$ & $\delta_{G}$ & $v_{G}$ & $H_{B}$ & $H_{E}$ & $m_{phot}$ & $z_{D}$      & $\lambda$      & $\phi$ \\
       & $[^{\circ}$] & $[^{\circ}$] & km.s$[^{-1}$] & [km] & [km] & [kg]       & $[^{\circ}$] & W $[^{\circ}$] & N $[^{\circ}$] \\
\hline
2 & 16.1 $\pm$ 2.0 & -24.5 $\pm$ 0.2 & 61.1 $\pm$ 0.6 & 138.8 & 89.9 & 1.8x10$^{-2}$ & 49.3 & 150.949 $\pm$ 0.016 & 18.247 $\pm$ 0.002 \\
5 & 16.0 $\pm$ 4.0 & -25.3 $\pm$ 0.9 & 61.5 $\pm$ 0.8 & 108.8 & 94.5 & 8.6x10$^{-4}$ & 50.6 & 151.240 $\pm$ 0.018 & 18.975 $\pm$ 0.004 \\
\hline
\end{tabular}
\begin{tablenotes}
\footnotesize
\item $\alpha_{G}$, $\delta_{G}$ ... geocentric radiant, $v_{G}$ ... geocentric velocity, $H_{B}$ ... beginning height, $H_{E}$ ... terminal height, $m_{phot}$ ... photometric mass, $z_{D}$ ... zenith distance of radiant, $\lambda$, $\phi$ ... geographic longitude W and latitude N of the beginning point.
\end{tablenotes}
\end{threeparttable}
\end{table*}

For all remaining single station meteors, the radiant of the brightest meteor was used for estimation of their atmospheric trajectories. The atmospheric trajectories for the entire cluster are shown in the right-hand panel of Figure~\ref{fig_3D_maps}.

The cluster occurred to the east of the Hawaii island. The brightest meteor, which was the second one observed, was firstly observed at the distance of 426 km from the Mauna Kea observatory. Gradually other meteors followed closer and closer to the island. The last one started 230 km from the station. 

As in the previous case, the pre-atmospheric positions of the meteoroids were also determined. The shortest distance measured between two fragments was 26 km, and the longest was 195 km. The total volume of the cluster was approximately 1.4~million km$^{3}$.

\subsection{Clusters origin}

Using a Monte Carlo method, we found that the probability of a random occurrence of such a compact cluster is extremely small. For a conservative estimate in case of MC/2023-04-20 cluster, we assumed that the camera is in operation 12 hours a day throughout the year. Meteor events have a uniform distribution with a frequency of 20 meteors per hour, which is almost twice the average hourly rate for sporadic meteors under a limiting magnitude of 6.5 \citep{Schmude1998}. In each such set (with 87\,600 random events, corresponding to the number of meteors recorded per year under the above-mentioned conditions) we searched for a 2.8-second time window that would contain 7 events. We repeated this test a million times. No positive result was detected. Only when we increased the frequency to 50 meteors per hour, one case occurs in a million tests. In the case of the MC/2024-07-11 cluster, where we test for the presence of a 3.8-second window containing 16 events, the probability of random occurrence is even smaller. In both cases, the null hypothesis that the cluster is a result of random coincidence can be rejected at a significance level well above $3\sigma$.

It appears that all clusters of meteoroids observed to date were formed by the breakup of a parent meteoroid a few days before its encounter with Earth. The trajectories of individual fragments are controlled by ejection velocities, solar radiation pressure (SRP), and, during the final phase, also by Earth's gravitational field and atmospheric drag. The SRP accelerates the fragments in an antisolar direction. This acceleration increases with decreasing fragment mass and decreasing bulk density. If there is enough time, the cluster evolves to a state where the less massive meteoroids are shifted away from the Sun relative to the more massive ones. For clusters arranged in this way, the method described in \cite{Capek2022} can be used to estimate not only the age of the cluster, but also the individual ejection velocities of the fragments. According to this method, it is possible (based on the knowledge of the mutual positions of $N$ fragments, their masses, and an assumed density) to derive $3N$ equations for $3N + 1$ unknowns, namely the ejection velocity vectors and the unknown age. A unique solution to this system does not generally exist; only the dependence of the ejection velocities on age can be determined. However, for well-arranged clusters, the age can be estimated from the minimum initial kinetic energy of all fragments. The reliability of this estimate, however, decreases with decreasing statistical significance of the cluster arrangement.

\begin{figure*}[htbp]
  \centering
  \includegraphics[width=0.9\textwidth]{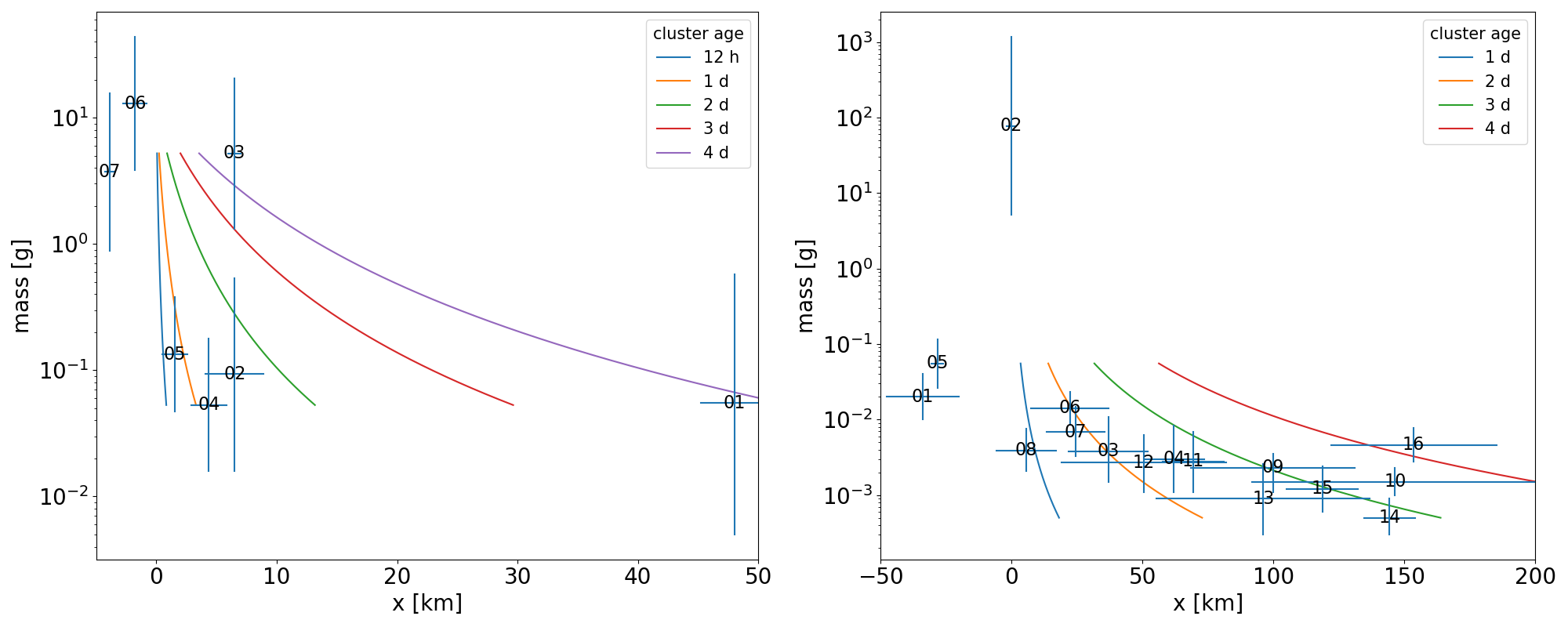}
  \caption{Mass of fragments vs. their positions in the anti-solar direction for 2023 (left plot) and 2024 (right plot) cluster. The colour lines shows expected positions with zero ejection velocity for different age of the cluster.}
  \label{fig_m_x}
\end{figure*}

\begin{figure}[htbp]
  \centering
  \includegraphics[width=0.49\textwidth]{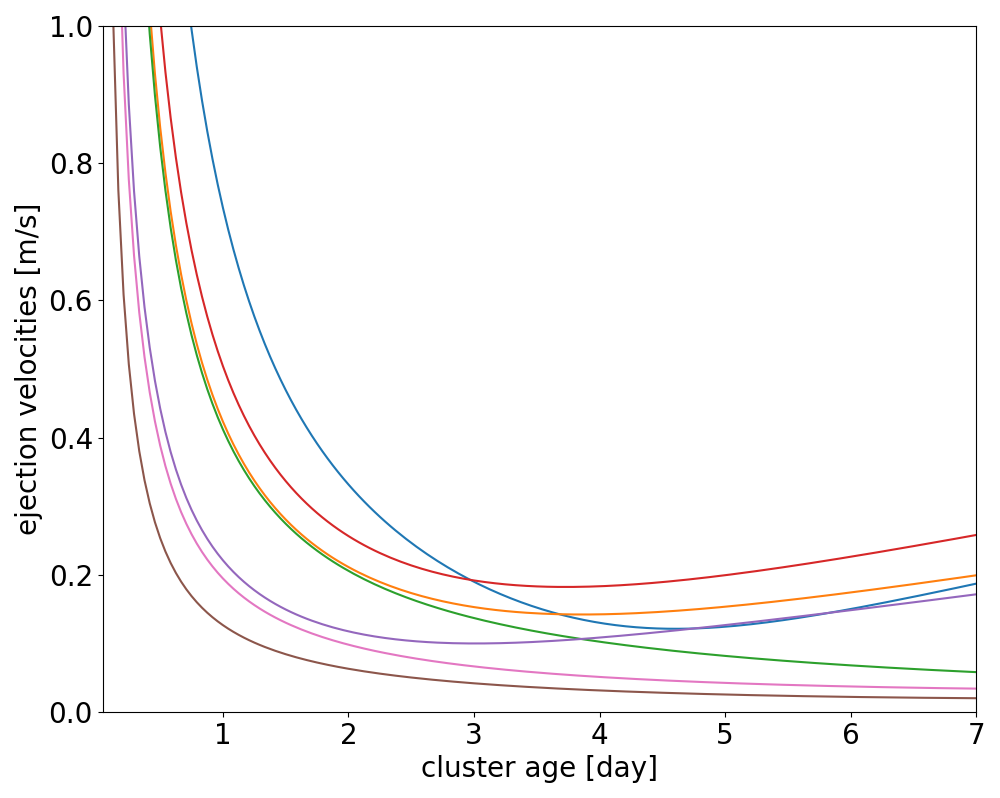}
  \includegraphics[width=0.49\textwidth]{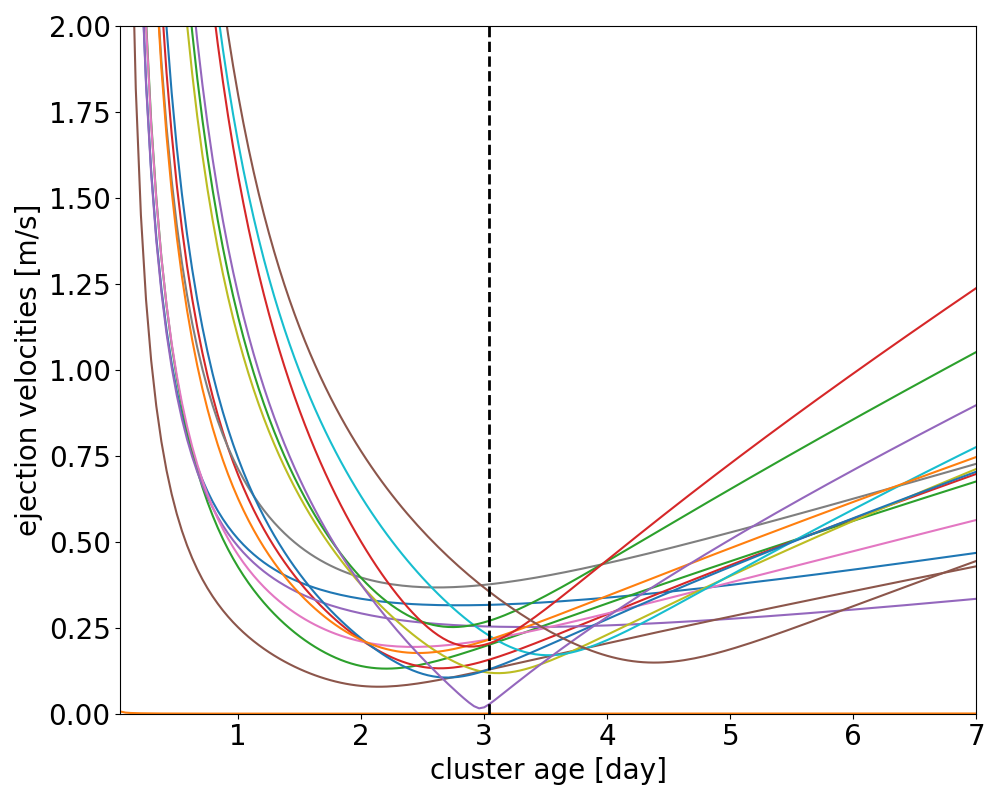}
  \caption{Ejection velocity vs. age of the cluster for 2023 (left plot) and 2024 (right plot) cluster. It was not possible to determine the age of 2023 cluster. In case of 2024 cluster, the age $\sim 3$~days is marked by dashed vertical line.}
  \label{fig_vej_x}
\end{figure}

In the case of the cluster from the 20th of April 2023, the correlation between mass and relative distance from the Sun (x-axis), which is shown in Figure~\ref{fig_m_x}, left, can be described by Spearman's correlation coefficient of $-0.54$. Although this is a moderately strong correlation, the $p$-value is $0.22$. This correlation is not statistically significant. Therefore, we cannot therefore rule out the possibility that this cluster arrangement is the result of chance. Figure~\ref{fig_vej_x}, left shows the dependence of ejection velocities on the age of this cluster. Using the method of \cite{Capek2022}, however, the age cannot be determined. For higher ages, the ejection velocities (especially for the least massive fragments) would be oriented in a narrow cone towards the Sun. On this basis, we estimate that the cluster is no older than 4~days. We are currently developing a method that would allow a more sophisticated estimate of the range of probable ages for similar clusters without mass separation. Regarding the magnitude of ejection velocities, at ages exceeding 18~hours, all ejection velocities are less than 1~m.s$^{-1}$. Such low ejection velocities imply that thermal stresses are the most likely cause of cluster formation. If we consider that the typical ejection velocity of a fragment ejected by the impact of an interplanetary dust particle is 10~m.s$^{-1}$, then the cluster would have to be less than 2 hours old, which cannot be completely ruled out.

In the cluster from the 11th of July 2024, we observe a very strong correlation between the mass and the position of the fragment on the x-axis (Figure~\ref{fig_m_x}, right); Spearman's coefficient is $-0.77$. In addition, the $p$-value is only $4.4\times 10^{-4}$, so at a significance level of $3 \sigma$, we can reject the hypothesis that this arrangement was created by chance. According to \cite{Capek2022}, the age of the cluster is 3~days and the ejection velocities range from approximately $0.03$~m.s$^{-1}$ to $0.4$~m.s$^{-1}$, see Figure~\ref{fig_vej_x}, right. Similarly to all clusters studied so far that show a statistically significant arrangement, these velocities are less than 1~m.s$^{-1}$. As in these cases, it is difficult to explain the disintegration of the parent meteoroid by the impact of interplanetary dust particles or by rotational fission. The most likely cause is the influence of thermal stresses. 

\section{Summary of meteor clusters}
\label{cluster_summary}

Up to now, ten meteor clusters have been analysed using the same methodology \citep{Capek2022}. The basic information on these clusters is summarised in Table~\ref{tab_cluster_overview}. Along with the new and the original designation, the table provides information on the date of the cluster's appearance, the number of the observed fragments, the cluster's duration, whether the fragments are aligned along the antisolar direction according to their masses, whether the cluster contains a dominant mass fragment, and the arrangement of the observation experiment. Single-station observations allow atmospheric trajectories to be estimated. The double- or multi-station observations allow more precise trajectory calculations. This designation means that at least one meteor was observed from at least two different stations. Note that even among these cases, there can be single-station meteors. Some of the cases in the table have already been published. This paper reports on the MC/2023-04-20 and MC/2024-07-11 clusters; the cases of the MC/2021-07-20 and MC/2024-08-15 clusters have yet to be analysed in detail.

The physical properties of the analysed clusters are shown in Table~\ref{tab_clusters_properties}. The volume of the cluster is calculated as the volume of a convex hull that encloses all meteoroids at the time of their atmospheric entry. This parameter demonstrates the compactness of the cluster, and there are truly extreme differences between the cases processed—from the most compact MC/2021-07-20 cluster (5.8~km$^3$) to the huge MC/2022-10-30 cluster (24~million~km$^3$).

\begin{figure}[htbp]
  \centering
  \includegraphics[width=0.48\textwidth]{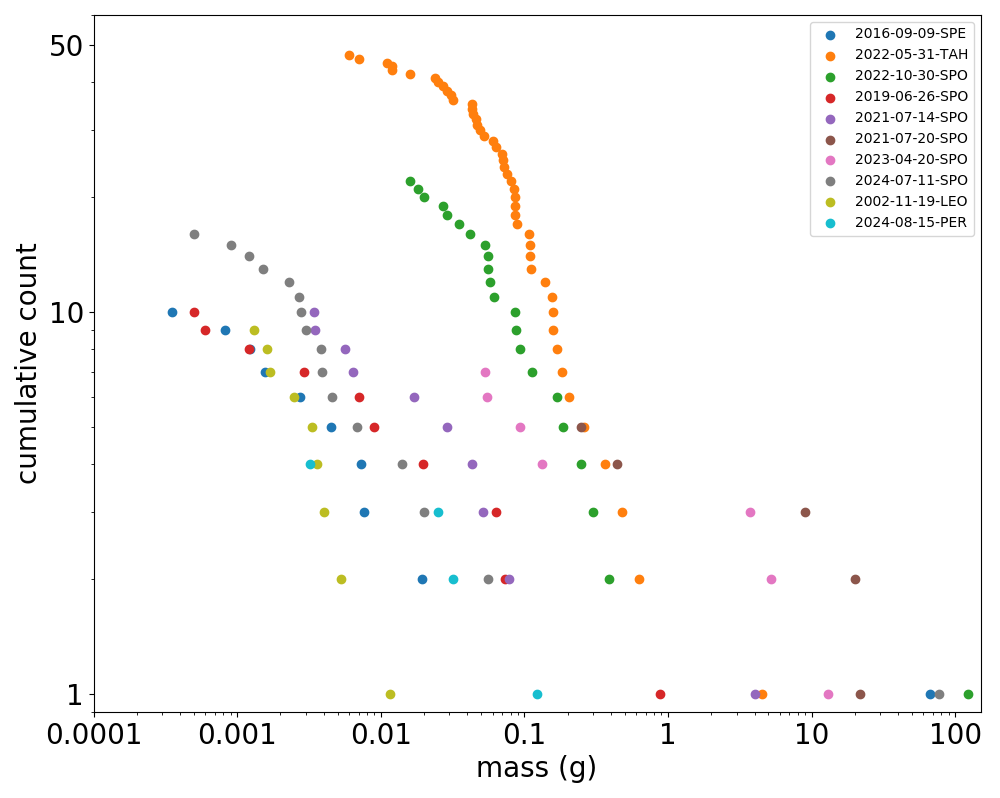}
  \caption{Cumulative distribution of fragment masses for all meteoroid clusters.}
  \label{fig_mass_cumul}
\end{figure}

The table also contains data on the total photometric mass of all observed fragments. We do not give the total mass of the cluster, i.e., including low-mass fragments that produce meteors too faint to be recorded. We have estimated this mass, or its range, in studies devoted to specific clusters. For this purpose, it is necessary to determine the mass distribution index, which can be a source of inaccuracies. The sophisticated \citet{Vida2020} method cannot be used directly. The mass distribution index probably differs for different mass intervals, as indicated by a comparison of Figure~\ref{fig_mass_cumul} with the cumulative mass distribution of fragments from the laboratory fragmentation experiment by \citet{Capek2025}.

An interesting quantity is the ratio of the mass of the largest member of the cluster to the total mass of all observed fragments. For clusters that show an arrangement of fragment masses according to their distance from the Sun, the most massive fragment represents from 81\% to more than 99.9\% and can be considered a remnant of the parent meteoroid. Clusters that do not have this arrangement have a significantly lower ratio, 33\%-67\%. 

Age and ejection velocities can be determined more accurately using the methodology of \citet{Capek2022} only for clusters that have the above-mentioned arrangement. This is due to the sufficiently long action of solar radiation pressure. For arranged clusters, the ages are between 1.3~days and 10.6~days. The corresponding ejection velocities are less than 1~m.s$^{-1}$ for all fragments. For unarranged clusters, it is only possible to roughly estimate the upper age limit, with the possibility of very low ages and thus higher ejection velocities. The table also provides information on the type of the meteoroid according to the Ceplecha's $K_{B}$ criterion \citep{Cep1988}. No uniform value was found for the fragments of each cluster. However, the spread of the values is rather similar to that typically observed in meteor showers. Further, the cluster's initial velocity when it entered the atmosphere and the most likely formation mechanism is provided.

\begin{table*}
\begin{threeparttable}
\caption{Basic information on the analysed meteor clusters.}
\label{tab_cluster_overview}
\centering          
\begin{tabular}{l l c c c c c c c}
\hline
Designation     & Original     & Date       & N        & $\Delta$T & Mass       & Dominant    & Observation     & Reference \\
                & name         &            &          & [s]       & separation & mass        &                 &           \\
\hline
MC/2002-11-19  &  LEO         & 19.11.2002 & 9        & 0.48      & no         & no          & single          & (1)       \\
MC/2016-09-09  & SPE          & 9.9.2016   & 10       & 1.5       & yes        & yes         & multi           & (2, 3)    \\
MC/2019-06-09  & Brazil       & 9.6.2019   & 10       & 0.5       & yes        & yes         & single          & (4)       \\
MC/2021-07-15  & Hawaii 2021  & 15.7.2021  & 15       & 10.9      & yes        & yes         & double          & (5)       \\
MC/2021-07-20  & Croatia      & 20.7.2021  & 5        & 1.15      & yes        & yes         & multi           & (11)      \\
MC/2022-05-31  & TAH          & 31.5.2022  & 51       & 8.5       & no         & no          & single          & (6, 7)    \\
MC/2022-10-30  & Scandic      & 30.10.2022 & 22       & 10        & yes        & yes         & multi           & (8)       \\
MC/2023-04-20  & Hawaii 2023  & 20.4.2023  & 7        & 2.8       & no         & no          & single          & (9, 10)   \\
MC/2024-07-11  & Hawaii 2024  & 11.7.2024  & 16       & 3.8       & yes        & yes         & double          & (9, 10)   \\
MC/2024-08-15  & PER          & 15.8.2024  &  4       & 1.3       & no         & yes         & double          & (11)      \\  
\hline
\end{tabular}

\begin{tablenotes}
\footnotesize
\item $Designation$ is a new and systematic name of the cluster, $Original~name$ is name of the cluster used in an original paper, $Date$ is a date of its appearance. $N$ denotes the number of members within the cluster; $\Delta$T is the time between the appearance of the first and the last meteor of the cluster; column $Observation$ provides information on number of stations the cluster was observed from.
\item References: (1) \cite{Koten2025b}, (2) \cite{Koten2017}, (3) \cite{Capek2022}, (4) \cite{Koten2024b}, (5) \cite{Abe2026}, (6) \cite{Vaubaillon2023}, (7) \cite{Koten2025}, (8) \cite{Koten2024}, (9) \cite{Watanabe2025}, (10) this work, (11) in preparation.
\end{tablenotes}
\end{threeparttable}
\end{table*}

\begin{table*}
\begin{threeparttable}
\caption{Physical properties of ten in detail analysed meteor clusters.}
\label{tab_clusters_properties}
\centering          
\begin{tabular}{l r c c c c c c c c}
\hline
               &  Volume           & Observed  & Main body  &    Age     & $v_{Ej}$       & Type  & $v_{\infty}$      & Thermal \\
               &  [1000 km$^{3}$]  & mass [g]  & mass [\%]  &   [days]   & [m.s$^{-1}$]   &       & [km.s$^{-1}$] & stress  \\
\hline                                                                                                          
MC/2002-11-19 &     0.93           &   0.035   & 33         &  <2–3      &  <1.0\tnote{1} &  C    & 72.2         & yes/?  \\
MC/2016-09-09 &    43.22           &  66.4     & >99.9      &   2.3      &  0.1–0.8       &  C    & 66.0         & yes    \\
MC/2019-06-09 &     0.22           &   1.049   & 83         &   1.1      & <0.4           &  A    & 63.5         & yes    \\
MC/2021-07-15 &  6545.62           &   4.249   & 94         &  4.6       & <0.5           &  C    & 51.6         & yes    \\
MC/2021-07-20 &   5.8~km$^3$       &  51.562   & 81 (2 pcs) &  1.3       & 0.01 - 0.03    &  C    & 27.8         & yes    \\
MC/2022-05-31 &   234.54           &   9.357   & 49         &  <3\tnote{$\dagger$} & <1.0\tnote{2}  &  D    & 16.9         & yes/?  \\
MC/2022-10-30 & 24627.82           & 124.911   & 98         &  10.6      & 0.1–0.6        &  C    & 68.0         & yes    \\
MC/2023-04-20 &    18.9            &  22.282   & 58         &  <4        & <1\tnote{2}    &  ast  & 44.0         & yes/?  \\
MC/2024-07-11 &   458.56           &  77.622   & 99.8       &  3         & 0.03 - 0.4     &  C    & 62.4         & yes    \\
MC/2024-08-15 &     0.18           &  0.181    & 67         &  1         & <1\tnote{3}    &  C    & 61.6         & yes/?  \\
\hline
\end{tabular}

\begin{tablenotes}
\footnotesize
\item $Volume$ the volume of the convex hull that encloses all meteoroids at time of entering the atmosphere, $Main~body~mass$ represents the percentage of the main body on the observed mass of the cluster, $Age$ is time from the formation of the cluster, $v_{Ej}$ is the ejection velocity, $Type$ is the meteoroid type, $v_{\infty}$ is the initial velocity, and $Thermal~stress$ indicates if the thermal stress was the most probable mechanism of cluster formation. In connection with $/?$ it shows that also other mechanisms cannot be excluded.
\item[$\dagger$] recent estimate based on directions of ejection velocities
\item[1] if age of the cluster is > 4 hours.
\item[2] if age of the cluster is > 1 day.
\item[3] if age of the cluster is > 6 hours.
\end{tablenotes}
\end{threeparttable}
\end{table*}

Although the sample size is still relatively small, certain trends are already emerging. Mass separation in the anti-solar direction increases with the age of the cluster and with the presence of the mass-dominant body. As we already know from previous studies, determining the age and the ejection velocities for such clusters is more reliable. However, there are four clusters for which mass separation is not observed. These include the MC/2002-11-19, MC/2022-05-31, MC/2023-04-20 and MC/2024-08-15 clusters. The main body of these clusters contributes about 60\% or less to the total mass of the cluster. Unsurprisingly, these clusters belong to the youngest ones. On the other side, there were also recorded clusters of similar age that have already separated according to the masses of their fragments. The lowest main body mass to total mass ratio for a cluster whose fragments have already separated is 78.5\% in the case of the MC/2021-07-15 cluster. It suggests that even the youngest clusters may be already mass separated if the mass distribution of the fragments is favourable.

From this point of view, the MC/2021-07-20 cluster observed in Croatia represents a specific case in which two main pieces contributing 81\% to the total mass together. As previously mentioned, this case is still being analysed in detail.

\begin{figure}[htbp]
  \centering
  \includegraphics[width=0.48\textwidth]{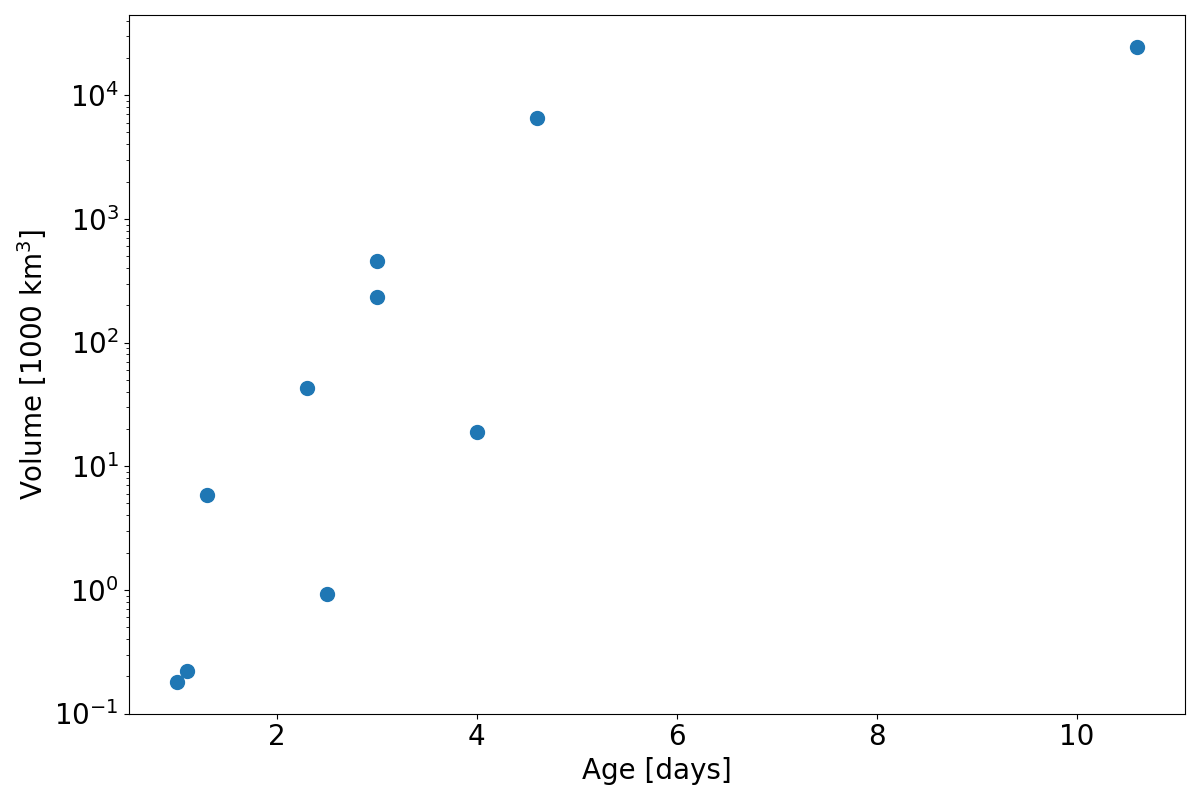}
  \caption{The volume of the cluster as a function of its age.}
  \label{fig_vol_age}
\end{figure}

All the clusters have been formed less than 11 days before their atmospheric encounter. It means that the fragmentation process has occurred relatively close to the Earth. The oldest recorded cluster, the MC/2022-10-30 cluster, was 10.6$\pm$1.7~days old. Taking into account its initial velocity, it was formed at the distance of 62$\pm$10 million kilometres from the point of the encounter with our planet. All other clusters were formed closer to the Earth. It is not an evidence that our planet influences the fragmentation of the meteoroids in its neighbourhood. It is rather a selection effect. As all the reported events have been recorded by a single station or by a multi station but still local experiment, only relatively close meteoroids in space were detected. Older clusters, if they existed, were not detected because they were already dispersed too much to be detected as clusters. Figure~\ref{fig_vol_age} shows that the volume of the cluster increases with an increasing age of the cluster.

The MC/2022-10-30 cluster was most likely a lucky event. It was only recorded as a cluster because its fragments were dispersed in the direction of the double station experiment in Norway. If it had been oriented differently, it would have avoided the detection as a cluster. There are two other clusters of a similar volume, but significantly younger. This suggest that the volume of the observed cluster is not only defined by its age. The size and direction of the ejection velocities of the fragments also contributes. Nevertheless, the trend of greater volume with higher age is evident. This also supports the idea that even older cluster are not observed because they have already dispersed too much to be detected as the clusters.

In all cases, thermal stress was found to be the most likely mechanism of cluster formation, although alternative scenarios cannot be excluded for cases with non-mass-separated fragments.  This mechanism is favoured because of low ejection velocities. Other scenarios including the collision in the interplanetary space and the rotation fission would produce higher ejection velocities. And again, cluster formed in this way may remain undetected because higher ejection velocities would cause quick dispersion of the fragments. 

Therefore, the fact that we only observed young meteor clusters formed close to the Earth by the thermal stress mechanism is merely a selection effect. To detect more dispersed clusters that were created far from the Earth and/or by a different mechanism, we need to simultaneously cover a larger volume of the atmosphere. Now is time for the global meteor networks.

The results of \cite{Ashimbekova2025} show promising results. 16 ''intercontinental'' cluster candidates have been identified. Nevertheless, a detailed statistical analysis must be carried out in order to distinguish between ordinary members of a meteor shower and real meteor clusters.

With only two exceptions, all the clusters originate from meteoroid of cometary origin. The tendency of meteoroids to break up due to thermal stresses, which are most likely responsible for the formation of arranged clusters, depends on several key parameters. Besides the (unknown) rotational state and thermophysical properties, the tensile strength of the parent meteoroid material is a crucial factor. Studies of the behaviour of larger meteoroids during atmospheric entry show that bodies of cometary origin are much more fragile and prone to fragmentation than bodies of asteroidal origin \citep[e.g.][]{Borovicka2022b}. Determining the specific value of tensile strength is only possible to an order of magnitude and with a high degree of uncertainty. It is commonly estimated from the dynamic pressure at the moment of major fragmentation in the atmosphere. For the most massive meteoroid from the MC/2016-09-09 cluster, which was about 5 cm in size, the tensile strength was estimated to be approximately $0.1$~MPa, while for the Scandinavian cluster, a tensile strength of approximately $0.06$~MPa was assumed. For meteoroids of asteroidal origin, we can expect tensile strengths up to an order of magnitude higher \citep{Henych2024}. We can therefore conclude that the material of cometary meteoroids can be disrupted more easily by thermal stresses than that of asteroidal meteoroids. Another factor may be the shape of the orbit, which affects the time during which the body is closer to the Sun and where the effects of thermal stress are more significant than in the vicinity of the aphelion. However, a more detailed analysis of this topic is beyond the scope of this article and will be discussed in more detail in a subsequent study.

\section{Conclusions}
\label{conclusions}
With an increase of the number of instrumentally detected and systematically described clusters, the opportunity arises to analyse their common properties in more detail. This paper therefore summarises the existing data on the already analysed clusters, together with details of two additional cases that have recently been recorded and added to the list. 

A suggestion is provided for how to name the clusters systematically. Furthermore, the conditions for distinguishing the clusters from the shower or sporadic background activity are suggested. Instead of strict limits on spatial and temporal proximity, statistical significance test and test of common physical origin are preferred.

The methods were demonstrated using two cases of clusters recorded by the video cameras located on the Hawaiian Islands. Clusters MC/2023-04-20 and MC/2024-07-11 illustrate two different scenarios. The first example is a cluster that shows no mass separation in the antisolar direction and contains no mass dominant body. Only an upper age limit can be determined. While thermal stress still remains likely formation scenario, other possible processes cannot be ruled out if higher ejection velocities are permitted. In the second case, both the mass dominant body and the mass separation were observed. Therefore, the age and the ejection velocities could be reliably determined, leading us to consider thermal stress as the most likely mechanism of the cluster formation. 

So far, all the cases studied have originated in the vicinity of the Earth. We believe this is a selection effect, given that the clusters were recorded by the local video experiments or local networks. These experiments only allow us to detect clusters with a relatively little dispersion of the fragments. Older clusters, i.e. clusters that originated far away from the Earth, or clusters that were formed by a different mechanism requiring higher ejection velocities, could escape detection simply because the mutual distances of the fragments would be greater than the detection area of the experiments used. There are the global meteor networks that can detect more dispersed clusters, which we call ''intercontinental'' clusters. Nevertheless, great care must be taken when dealing with such global data. With more dispersed fragments, the condition on their proximity may not be met. In particular, it could be difficult to distinguish potential clusters from the ''normal'' shower activity during the more active meteor showers. It seems that the ''intercontinental'' clusters will be easier to identify in the case of sporadic ones.

With an increasing number of the meteor cameras around the world, we can expect clusters to be detected more frequently. This paper may also serve as a guide to other authors on how to analyse data consistently. 

\section{Acknowledgements}
Work of PK and D\v{C} was supported by the Grant Agency of the Czech Republic grants 20-10907S and 24-10143S, and the institutional project RVO:67985815. JT and TV weres supported by the Slovak Research and Development Agency grant APVV-23-0323 and VV-MVP-24-0232.

\bibliographystyle{cas-model2-names}
\bibliography{refs.bib}

\end{document}